\documentstyle{article}
\title{Conservation laws and angular transverse shifts of
the reflected and transmitted light beams}
\author{V.G.Fedoseyev\\
Institute of Physics, University of Tartu,\\
Riia 142, 51014 Tartu, Estonia\\
E-mail address: fedo@fi.tartu.ee\\
Fax: +372-7383033}
\date{\it June 20, 2008}
\begin{document}
\maketitle
\begin{abstract}
\indent

The relation between the angular transverse shifts of the beams
reflected and transmitted at a plane interface of two isotropic
transparent media is established. The evaluation of this relation
is based on the conservation law of the transverse component of
the Minkowski linear momentum, which takes place in the processes
of the reflection and transmission of wave packets.
\end{abstract}



\section*{I. Introduction}
\indent

It is known that in some cases light beams undergo transverse
shifts (TSs) after reflection and transmission at a plane
interface of isotropic transparent media. Two types of TSs are
known: the linear TS (LTS) and the angular TS (ATS).

The LTS phenomenon means that the axis of the secondary beam is
parallel to the plane of incidence, but it does not lie in it.
This effect is connected with the transformation of the spin or
intrinsic orbital angular momentum at reflection and transmission
of the wave packet. The spin-dependent LTS has been studied for
more than half a century (see, for instance, [1-13] and the
references therein). In the case of total reflection, this effect
is usually called the Imbert-Fedorov shift. The LTSs of the
reflected and transmitted beam, which depend on the intrinsic
orbital angular momentum of the incident beam, have been predicted
in [14, 15]; the experimental investigations of this and the
associated effects have been carried out in [16-18].

The ATS phenomenon means that the actual axis of the secondary
beam is inclined to the plane of incidence, the value of the ATS
is given by the angle of inclination. Like in the case of LTSs,
two types of ATSs are known: the spin-dependent and the
spin-independent ATSs. The expression for the spin-dependent ATS
of the reflected beam has been obtained in the work by Nasalski
[9]. Bliokh and Bliokh [13] have performed explicit calculations
of spin-dependent ATSs for both reflected and transmitted beams.
It should also be mentioned that the possibility of these ATSs has
been pointed out in [7]. Spin-independent ATSs of the reflected
and transmitted beams have been predicted in Ref. [14] and in the
papers by Alda [19] and by Alda and Rico-Garcia [20].

The LTSs of the secondary beams are controlled by the conservation
law of the normal-to-interface component of the Minkowski angular
momentum [21-24], what is explained as follows. If the normal
component of the total intrinsic angular momentum changes after
reflection and transmission, the above-mentioned law demands that
the centres of gravity of the secondary packets must undergo
shifts in the transverse direction in order that the normal
component of the total extrinsic AM generated by the shifts could
compensate this change. In application to the totally reflected
beam, this fact was known long ago [3]. As for the LTSs of
partially reflected and transmitted beams, these phenomena have
been especially intensively discussed from the point of view of
the conservation laws in the last years [12, 13, 24].

In this communication, it will be shown that the ATSs of the
secondary beams are also controlled by a conservation law, namely
by conservation of the transverse, i.e. perpendicular to the plane
of incidence, component of the Minkowski linear momentum.

\section*{II. Geometry of reflection and transmission}
\indent

 Let us consider the reflection and transmission of a monochromatic
light beam at a plane interface of two semi-infinite transparent,
isotropic, nondispersive, and nonmagnetic media. The scheme of
this process is shown in Fig. 1. The position of the interface is
defined by the equation $\hat{\bf N}\cdot {\bf x}=0$, where ${\bf
x}$ is the 3D radius vector, and $\hat{\bf N}$ is the unit normal
to the interface directed from the first medium, which fills the
upper half-space, to the second one. The beam is assumed to be
incident from the first medium. Let us denote the refractive
indices of the media in the upper half-space ($\hat{\bf N}\cdot
{\bf x}<0$) and the lower half-space ($\hat{\bf N}\cdot {\bf
x}>0$)  by $n\sp{(1)}$ and $n\sp{(2)}$, respectively.

Throughout this paper, we shall use the superscripts $i$, $\rho$,
and $\tau$ for indicating the quantities characteristic of the
incident, reflected and transmitted beams, respectively. The
superscript $a$ will be used in order to designate an arbitrary
(incident, reflected or transmitted) beam and the superscript $s$
in order to designate one of the secondary beams (reflected or
transmitted). Thus, $a$=$i$, $\rho$, or $\tau$, and $s$=$\rho$ or
$\tau$. By $n\sp{(a)}$ the refractive index of the medium will be
denoted, in which the $a$-th beam propagates, i.e.
$n\sp{(i)}=n\sp{(\rho )}=n\sp{(1)}$, and $n\sp{(\tau
)}=n\sp{(2)}$.

We shall employ four coordinate systems. Three systems are
connected with the incident, reflected and transmitted beams.
These systems will be termed the $i$-, $\rho$, and $\tau$-systems,
their bases are given by the unit vectors $\hat{\bf x}_1\sp{(a)}$,
$\hat{\bf x}_2\sp{(a)}$, and $\hat{\bf x}_3\sp{(a)}$. In the
$i$-system, the $x_3\sp{(i)}$ axis is assumed to coincide with the
incident beam's axis (the exact definition of the latter will be
given later). This axis and the unit vector $\hat{\bf N}$ define
the position of the beam's plane of incidence in 3D space; the
angle between the unit vectors $\hat{\bf N}$ and $\hat{\bf
x}_3\sp{(i)}$ is the beam's angle of incidence: $\theta\sp{(i)}=
\arccos (\hat{\bf N}\cdot \hat{\bf x}_3\sp{(i)})$. The coordinate
origin $O$ in every system is taken to be the point of
intersection of the incident beam axis with the interface. In the
$\rho$- and $\tau$-systems, the $x_3\sp{(\rho)}$ and
$x_3\sp{(\tau)}$ axes are assumed to coincide with the
geometric-optical axes of the reflected and transmitted beams. The
geometric-optical axes of the secondary beams are defined as the
rays which are intersected by the interface at the coordinate
origin and whose directions are characterized by the unit vectors
$\hat{\bf x}_3\sp{(\rho)}$ and $\hat{\bf x}_3\sp{(\tau)}$, these
vectors being related to $\hat{\bf x}_3\sp{(i)}$ through the Snell
laws. The beam's angles of reflection and transmission
$\theta\sp{(\rho )}$ and $\theta\sp{(\tau )}$ are the angles
between the vector $\hat{\bf N}$ and the respective geometric
optics axis. $\theta\sp{(\rho )}=\pi-\theta\sp{(i)}$, while
$\theta\sp{(\tau )}$ is defined by the relation
$n\sp{(2)}\sin\theta\sp{(\tau)}=n\sp{(1)}\sin\theta\sp{(i)}$.

Let us define the $x_2$ axis of every system to be perpendicular
to the plane of incidence, this axis (the transverse one) is
characterized by the unit vector $\hat{\bf x}_2=\hat{\bf N} \times
\hat{\bf x}_3\sp{(i)} / |\hat{\bf N} \times \hat{\bf
x}_3\sp{(i)}|$. The $x_1\sp{(i)}$, $x_1\sp{(\rho)}$, and
$x_1\sp{(\tau)}$ axes lie in the plane of incidence, the direction
of the $x_1\sp{(a)}$ axis is given by the unit vector $\hat{\bf
x}_1\sp{(a)} =\hat{\bf x}_2 \times \hat{\bf x}_3\sp{(a)}$.

In the $a$-th system, the 3D radius-vector is represented as
follows:
\begin{equation}
{\bf x}={\bf x}_{\perp}\sp{(a)}+x_3\sp{(a)}\hat{\bf x}_3\sp{(a)}\;
,
\end{equation}
where ${\bf x}_{\perp}\sp{(i)}$,  ${\bf x}_{\perp}\sp{(\rho)}$,
and ${\bf x}_{\perp}\sp{(\tau)}$ are the 2D planar radius vectors
lying in the planes which are perpendicular to the axis of the
incident beam or to the geometric optics axes of the respective
secondary beam:
\begin{equation}
{\bf x}_{\perp}\sp{(a)}=x_1\sp{(a)}\hat{\bf
x}_1\sp{(a)}+x_2\hat{\bf x}_2 \; .
\end{equation}

The fourth system is connected with the interface, its basis being
given by the unit vectors $\hat{\bf x}_1=-\hat{\bf N}$, $\hat{\bf
x}_2$, and $\hat{\bf x}_3=\hat{\bf x}_1\times\hat{\bf x}_2$.

\section*{III. Relation between the ATSs of the reflected and
transmitted beams}
\indent

We shall assume that the incident beam is paraxial, i.e. that
$\lambda\ll b$ , where $\lambda$ is the wavelength of the light in
vacuum and $b$ is a mean dimension of the beam in the planar
plane.

Let us denote the electric and magnetic field vectors of the
$a$-th beam at the point $\bf x$ by ${\bf E}\sp{(a)}({\bf x})$ and
${\bf H}\sp{(a)}({\bf x})$. The vector ${\bf F}\sp{(a)}({\bf x})$
will designate either ${\bf E}\sp{(a)}({\bf x})$ or ${\bf
H}\sp{(a)}({\bf x})$. The time dependence of the field vectors and
of the other characteristics of the beams is suppressed.

In order to consider the ATS phenomenon, it is necessary to define
the actual axes of the secondary beams. Let us first define the
centre of gravity of the $a$-th beam. Its position on the beam's
cross-section is given by the 2D vector
\begin{equation}
{\bf R}_{\perp}\sp{(a)}(X_3\sp{(a)})=\frac{\int {\bf
x}_{\perp}\sp{(a)}w\sp{(a)}({\bf
x}_{\perp}\sp{(a)},X_3\sp{(a)})d{\bf x}_{\perp}\sp{(a)}}{\int
w\sp{(a)}({\bf x}_{\perp}\sp{(a)},X_3\sp{(a)})d{\bf
x}_{\perp}\sp{(a)}}\; ,
\end{equation}
where $X_3\sp{(a)}$ is the $x_3\sp{(a)}$-coordinate of the
cross-section, and $w\sp{(a)}({\bf x})$ is the electromagnetic
energy density inside the $a$-th beam,
\begin{equation}
w\sp{(a)}({\bf x})=\frac{1}{8\pi}\left [(n\sp{(a)})\sp{2}\left
({\bf E}\sp{(a)}({\bf x})\right ) \sp 2+\left ({\bf
H}\sp{(a)}({\bf x})\right )\sp 2\right ] \; .
\end{equation}
ATSs are actual in the far-field region. In view of that, let us
assume that every $X_3\sp{(a)}$ satisfies the condition:
\begin{equation}
\vert X_3\sp{(a)}\vert >b\sp{2}/\lambda\; .
\end{equation}
Let us also assume that, if $\theta\sp{(a)}$ is close to $0$ or to
$\pi/2$, $X_3\sp{(a)}$ satisfy the additional condition:
\begin{equation}
\vert X_3\sp{(a)}\vert \gg b/\sin(2\theta\sp{(a)})\; .
\end{equation}
If the relation (6) is fulfilled, the electromagnetic fields of
the beams are negligible at the lines of intersection of the
respective cross-sections and the interface. Again, the reflected
field can be ignored on the cross-section of the incident beam and
vice versa.

In the regions (5), the geometrical loci of the centres of gravity
of the incident, reflected and transmitted beams are the straight
lines. Let us define the first one to be the axis of the incident
beam. The second and the third lines determine the actual axes of
the secondary beams. These axes and the unit normal $\hat{\bf N}$
give the actual planes of reflection and transmission, the angles
between them and the incidence plane will be denoted by
$\Phi\sp{(\rho )}$ and $\Phi\sp{(\tau )}$, respectively. Let us
define the ATSs of the reflected and transmitted beams,
$\phi\sp{(\rho )}$ and $\phi\sp{(\tau )}$, as the angles of the
inclination of the beams' actual axes to the incidence plane. The
angles $\phi\sp{(s)}$ and $\Phi\sp{(s)}$ are small [13, 14, 19,
20], thus, $\sin\phi\sp{(s)}\simeq\phi\sp{(s)}$ and
$\sin\Phi\sp{(s)}\simeq\Phi\sp{(s)}$. In this case, the relation
between $\phi\sp{(s)}$ and $\Phi\sp{(s)}$ is as follows:
\begin{equation}
\Phi\sp{(s)}\simeq\frac{\phi\sp{(s)}} {\sin\theta\sp{(s)}}\; .
\end{equation}
The meanings of the angles $\phi\sp{(\rho )}$ and $\Phi\sp{(\rho
)}$ are illustrated in Fig. 2, the meanings of the angles
$\phi\sp{(\tau )}$ and $\Phi\sp{(\tau )}$ are similar.

Let us now consider the ATS phenomenon from the point of view of
the dynamics of the processes of the reflection and transmission
of wave packets. In order to do that, let us fix a time instant
$t_0=n\sp{(1)}X_3\sp{(i)}$ ($t_0<0$) and select at this instant a
section of the incident beam, which is restricted with two
cross-sections, the axial coordinate of its centre being equal to
$X_3\sp{(i)}$ (see Fig. 3). Its length, i.e. the distance between
the cross-sections, will be denoted by $D$. Let us impose the
following restrictions on $D$:
\begin{equation}
\vert X_3\sp{(i)}\vert\gg D\gg\sqrt{\lambda \vert
X_3\sp{(i)}\vert} \; .
\end{equation}

Being separated from the beam, the selected section represents a
packet of electromagnetic waves. Qualitatively, the process of
propagation of the packet constructed in such a way is Fig. 3. At
the instant $t_0$ and during some interval after it, the influence
of the second medium on the incident field is negligible. Thus,
during this interval, the motion of the packet is quasi-free, i.e.
it occurs as if the packet propagates in the homogeneous medium
with the refracted index $n\sp{(1)}$. At $\vert t\vert\sim
n\sp{(1)}D$, the incident, reflected and transmitted fields
coexist. Finally, when $t$ is positive and large enough, the
reflected and transmitted packets are well formed, and they
propagate quasi-freely in the first and the second medium,
respectively. Further on, the time instants, when the motions of
the packets are quasi-free, will be specified by means of the
accent "tilde" over $t$, these times obey the condition:
\begin{equation} \vert\tilde t\vert\gg n\sp{(1)}D\; .
\end{equation}

Let us now perform a detailed analysis of the motion of the
constructed packet. Further on, the same letters will be used for
notations of field vectors and other characteristics of the $a$-th
beam and of the $a$-th packet. However, in order to make the
difference between the characteristics of the beam and of the
packet, in the latter case the respective letter will be
overlined. Again, when necessary, the dependence of the respective
characteristics on $t$ will be explicitly pointed out.

Our analysis of the motion of the constructed packet will be based
on the local energy-momentum conservation laws, which take place
at an arbitrary $t>t_0$. These laws can be written by means of the
Minkowski energy-momentum tensor $\overline{T}_{jk}({\bf{x}};t)$
[22, 25, 26]. In order to do this, let us introduce the conception
of a total electromagnetic field, whose characteristics will be
written down without superscript $a$. It is evident that
\begin{equation}
\overline{{\bf F}}({\bf x};t)=\overline{{\bf F}}\sp{(i)}({\bf
x};t)+\overline{{\bf F}}\sp{(\rho)}({\bf x};t),\;\;{\rm if}\;
x_1>0\; ,
\end{equation}
and
\begin{equation}
\overline{{\bf F}}({\bf x};t)=\overline{{\bf F}}\sp{(\tau )}({\bf
x};t),\;\; {\rm if}\; x_1<0\; .
\end{equation}
Again, let us denote by $\varepsilon (x_1)$ the dielectric
constant at an arbitrary point in the whole space: $\varepsilon
(x_1)=( n\sp{(1)})\sp{2}$, if $x_1>0$, and $\varepsilon (x_1)=(
n\sp{(2)})\sp{2}$, if $x_1<0$. Then, in the coordinate system
connected with the interface, $\overline{T}_{jk}({\bf{x}};t)$ will
look as follows:
\begin{equation}
\overline{T}_{jk}({\bf{x}};t)= \left(
\begin{array}{ll}
\overline{\sigma}_{\alpha \beta}({\bf{x}};t)  &-i{\overline{\bf{g}}}({\bf{x}};t)\\
-i\overline{\bf{s}}({\bf{x}};t) & {\overline w}({\bf{x}};t)
\end{array}
\right) \; ,
\end{equation}
where $j,k=1,2,3,4$, and $\alpha ,\beta =1,2,3$. The constituents
of this tensor are as follows. ${\overline w}({\bf{x}};t)$ is the
electromagnetic energy density at an arbitrary point and at an
arbitrary instant $t$:
\begin{equation}
{\overline w}\sp{(a)}({\bf x})=\frac{1}{8\pi}\left (\varepsilon
(x_1)\left ({\overline{\bf E}}\sp{(a)}({\bf x};t)\right ) \sp
2+\left ({\overline{\bf H}}\sp{(a)}({\bf x};t)\right )\sp 2\right
] \; .
\end{equation}
${\overline{\bf s}}({\bf x};t)$ is the Poynting vector:
\begin{equation}
{\overline{\bf s}}({\bf x};t)=\frac{1}{4\pi}{\overline{\bf
E}}({\bf x};t)\times{\overline{\bf H}}({\bf x};t) \; .
\end{equation}
We use the system of units, in which the light velocity in vacuum
is unity, in such a system $\overline{{\bf s}}\sp{(a)}({\bf x};t)$
coincides with the density of the linear momentum defined in the
sense of Abraham [22, 25, 26]. ${\overline{\bf g}}\sp{(a)}({\bf
x};t)$ is the density of the linear momentum defined in the sense
of Minkowski [22, 25, 26], in the employed system of units,
\begin{equation}
{\overline{\bf g}}({\bf x};t)=\varepsilon (x_1){\overline{\bf
s}}({\bf x};t) \; .
\end{equation}
$\overline{\sigma}_{\alpha \beta}(\bf{x};t)$ is the Maxwellian
stress tensor:
\begin{equation} \overline{\sigma}_{\alpha
\beta}({\bf{x}};t)=\frac{1}{4\pi}\left (\varepsilon
(x_1){\overline E}_{\alpha}({\bf{x}};t){\overline
E}_{\beta}({\bf{x}};t)+{\overline
H}_{\alpha}({\bf{x}};t){\overline H}_{\beta}({\bf{x}};t)-
\delta_{\alpha \beta}{\overline w}({\bf{x}};t)\right )\; ,
\end{equation}
were $\delta_{\alpha \beta}$ is the Kronecker symbol.

By means of the tensor $\overline{T}_{jk}({\bf{x}};t)$ the local
conservation laws of the energy ($j=4$) and of the components of
the Minkowski linear momentum ($j=1,2,3$) are written as follows:
\begin{equation}
\frac{\partial{\overline{T}_{j4}({\bf{x}};t)}}{\partial
t}=-i\sum^3_{\beta=1}\frac{\partial{\overline{T}_{j\beta}({\bf{x}};t)}}
{\partial{x_\beta}} \; ,
\end{equation}
Eq. (17) is correct in the whole space except the interface and at
an arbitrary instant of time. This equation can be used as a basis
for the description of the motion of the constructed packet at any
$t$.

Let us first consider quasi-free regimes. In these regimes, only
one packet exits in the first medium, i.e., in this medium,
${\overline{\bf F}}({\bf{x}};\tilde t)\cong {\overline{\bf
F}}\sp{(i)}({\bf{x}};\tilde t)$ when $\tilde t<0$, and
${\overline{\bf F}}({\bf{x}};\tilde t)\cong {\overline{\bf
F}}\sp{(\rho)}({\bf{x}};\tilde t)$ when $\tilde t>0$. Let us take
this fact into account and integrate Eq. (17) over the upper
half-space if $\tilde t<0$ and over the upper and lower
half-spaces separately if $\tilde t>0$. At that let us take into
account that ${\overline{\bf F}}\sp{(a)}({\bf{x}};\tilde t)$ is
negligible on the interface; thus, the integration over the
respective half-space may be replaced by the integration over the
whole space, and, as a consequence, the partial derivative
$\partial/\partial \tilde t$ in the left-hand side of the obtained
expression may be replaced by the full one $d/d\tilde t$. Then we
obtain that the electromagnetic energy
$\overline{W}\sp{(a)}(\tilde t)$ and the Minkowski linear momentum
$\overline{{\bf G}}\sp{(a)}(\tilde t)$ of the $a$-th packet do not
depend on $\tilde t$, i.e.
\begin{equation}
\overline{W}\sp{(a)}(\tilde t)=\int \overline{w}\sp{(a)}({\bf
x};\tilde t)d{\bf x}\cong \overline{W}\sp{(a)}\; ,
\end{equation}
and
\begin{equation}
\overline{{\bf G}}\sp{(a)}(\tilde t)=\int \overline{{\bf
g}}\sp{(a)}({\bf x};\tilde t)d{\bf x}\cong \overline{{\bf
G}}\sp{(a)}\; ,
\end{equation}
where $\overline{W}\sp{(a)}$ and $\overline{{\bf G}}\sp{(a)}$ are
the constants. Again, by using the relation (15) we obtain that
the linear momenta of the $a$-th packet defined in the senses of
Abraham and Minkowski are related by
\begin{equation}
\overline{{\bf S}}\sp{(a)}(\tilde t)=\int \overline{{\bf
s}}\sp{(a)}({\bf x};\tilde t)d{\bf x}\cong \overline{{\bf
S}}\sp{(a)}=\frac{\overline{{\bf G}}\sp{(a)}}{ (n\sp{(a)}
)\sp{2}}\; .
\end{equation}

Let us now consider the 3D radius-vector of the centre of gravity
of the $a$-th packet $\overline{{\bf R}}\sp{(a)}(\tilde t)$. It is
defined as follows:
\begin{equation}
\overline{{\bf R}}\sp{(a)}(\tilde t)=\frac{\int{\bf
x}\overline{w}\sp{(a)}({\bf x};\tilde t)d{\bf
x}}{\overline{W}\sp{(a)}}\; ,
\end{equation}
the integration in the right-hand side of Eq. (21) being performed
over the respective half-space. Let us act by the operator
$\partial/\partial \tilde t$ on the left-hand and right-hand sides
of Eq. (21). In the right-hand side of the obtained expression,
let use the relation (17) with $j=4$, and, by the calculation of
the numerator, let us perform the integration by parts. In the
left-hand side, the partial derivative $\partial/\partial \tilde
t$ may be replaced, due to the above reasoning, by the full one
$d/d\tilde t$. After carrying out these operations, we obtain the
following equation of the motion of the $a$-packet:
\begin{equation}
\frac{d\overline{{\bf R}}\sp{(a)}(\tilde t)}{d\tilde
t}=\frac{\overline{{\bf S}}\sp{(a)}}{\overline{W}\sp{(a)}}\; .
\end{equation}

Thus, the motion of the $a$-th packet under condition (9) is
rectilinear. Let us impose the following restriction on $\tilde
t$:
\begin{equation}
\tilde t\ll D\sp{2}/\lambda\; .
\end{equation}
If $\tilde t$ satisfies the relation (23), the vector
${\overline{\bf F}}\sp{(a)}({\bf{x}};\tilde t)$ approximately
coincides with the vector ${\bf F}\sp{(a)}({\bf{x}})$ inside the
section of $a$-th beam, whose centre is situated at $X_3\sp{(a)}=
\tilde t/n\sp{(a)}$ and whose length equals to
$Dn\sp{(1)}/n\sp{(a)}$. As a consequence,
\begin{equation}
\overline{{\bf R}}_{\perp}\sp{(a)}(\tilde t)\simeq {\bf
R}_{\perp}\sp{(a)}(\tilde t/n\sp{(a)})\; ,
\end{equation}
what means that, in the quasi-free regime, the centre of gravity
of the $i$-th packet moves along the axis of the incident beam,
while the centre of gravity of $s$-th packet moves along the
$s$-th actual axis.

Thus, the angular shifts $\phi\sp{(\rho)}$ and $\phi\sp{(\tau)}$
are equal to the angles of inclination of the vectors
$\overline{{\bf S}}\sp{(\rho )}$ and $\overline{{\bf S}}\sp{(\tau
)}$ to the incidence plane. Consider the properties of these
vectors. Let us begin with the zero-order approximation with
respect to $\lambda /b$. In this approximation, the laws of
reflection and transmission for the packet under consideration and
for the plane wave, which is incident at the angle
$\theta\sp{(i)}$ and whose polarization vector is $\hat{\bf
e}\sp{(i)}=\overline{{\bf E}}\sp{(i)}({\bf
x};t_0)/\vert\overline{{\bf E}}\sp{(i)}({\bf x};t_0)\vert$, are
equivalent. Let us denote the characteristics of the secondary
beams calculated in the zero-order approximation by means of the
additional superscript $0$. Then we have:
${\overline{W}\sp{(s0)}}=Q\sp{(s)}{\overline{W}\sp{(i)}}$, were
$Q\sp{(\rho )}$ and $Q\sp{(\tau )}$ are the reflectivity and
transmissivity of the above-mentioned plane wave [25, 27]. As for
the vectors $\overline{{\bf S}}\sp{(\rho 0)}$ and $\overline{{\bf
S}}\sp{(\tau 0)}$, they are directed along the geometric optics
axes of the reflected and transmitted beams, respectively:
$\overline{{\bf S}}\sp{(s0)}=\overline{S}_3\sp{(s0)}\hat{\bf
x}_3\sp{(s)}$. The values ${\overline{S}_3\sp{(s0)}}$ are related
to ${\overline{W}\sp{(s0)}}$ by ${\overline
S}_3\sp{(s0)}={\overline W}\sp{(s0)}/n\sp{(s)}$. The $x_1\sp{(s)}$
and $x_2$ components of the vector $\overline{{\bf S}}\sp{(s)}$
are of a highest order. In the nearest-order approximation, the
former does not affect the angle of inclination of the vector
$\overline{{\bf S}}\sp{(s)}$ to the incidence plane. Thus, by
using the above relations one obtains:
\begin{equation}
\phi\sp{(s)}\cong\frac{{\overline S}_2\sp{(s)}}
{S_3\sp{(s0)}}\cong\frac{n\sp{(s)}{\overline S}_2\sp{(s)}}
{Q\sp{(s)}{\overline{W}\sp{(i)}}}\; .
\end{equation}

Eq. (25) shows that from the point of view of the dynamics the
calculation of the ATS of the secondary beam $\phi\sp{(s)}$ turns
to the calculation of the transverse linear momentum ${\overline
S}_2\sp{(s)}$, which is obtained by the packet after the
reflection or transmission. In order to carry out these
calculations, it is necessary to make an assumption about the
structure of the incident beam. However, a relation between
$\phi\sp{(\rho )}$ and $\phi\sp{(\tau )}$ can be obtained without
any assumption. The derivation of this relation is carried out on
the basis of the conservation law of the transverse component of
the linear momentum defined in the sense of Minkowski. In order to
do that, let us take $j=2$ in Eq. (17) and perform the integration
of the left-hand and right-hand sides of this equation over the
whole space ($t$ is assumed to be arbitrary). By means of the
Gauss theorem, the 3D right-hand integral transforms into two 2D
integrals of $\overline{\sigma}_{21}({\bf x};t)$ over both sides
of the interface. By taking into account the boundary conditions
at the interface [25, 27], one can be convinced that
$\overline{\sigma}_{21}({\bf x};t)$ is continuous at the
interface; as a consequence, the sum of the two above-mentioned
surface integrals equals zero, therefore $d{\overline
G}_2(t)/dt=0$. At the instant $t_0$, ${\overline
G}_2(t_0)\cong{\overline G}_2\sp{(i)}(t_0)=0$, hence, ${\overline
G}_2(t)=0$ at an arbitrary $t$. Let us now consider the secondary
packets at a instant $\tilde t>0$, when they are well-formed. At
such an instant, ${\overline G}_2(\tilde t)={\overline
G}_2\sp{(\rho )}(\tilde t)+{\overline G}_2\sp{(\tau )}(\tilde t)$.
Hence, taking into account Eq. (19), we get: ${\overline
G}_2\sp{(\rho )}+{\overline G}_2\sp{(\tau )}=0$. By using in this
expression the relation between the quantities ${\overline
G}_2\sp{(a)}$ and ${\overline S}_2\sp{(a)}$ given by Eq. (20), one
obtains:
\begin{equation}
{(n\sp{(1)})\sp{2}}{\overline S}_2\sp{(\rho )}+{ (n\sp{(2)}
)\sp{2}}{\overline S}_2\sp{(\tau )}=0\; .
\end{equation}
Thus, the reflected and transmitted beams separately can obtain
transverse Abraham linear momenta, however, their values must obey
the condition (26).

Substitution of the expression for ${\overline S}_2\sp{(\rho ,\tau
)}$ obtained from Eq. (25) into Eq. (26) and the using of the
relation (7) yields
\begin{equation}
Q\sp{(\rho )}\Phi\sp{(\rho )}+Q\sp{(\tau )}\Phi\sp{(\tau )}=0\; .
\end{equation}
The relation between the ATSs of the reflected and transmitted
beams, $\phi\sp{(\rho )}$ and $\phi\sp{(\tau )}$, can be obtained
from Eq. (27) by means of Eq. (7). It should be underlined that
the relation (27) does not depend on the incident beam's
structure. Nevertheless, one can be convinced that the
spin-dependent and the spin-independent ATSs, which have been
calculated in Refs. [13] and [14, 19, 20], respectively, satisfy
the relation (27), although the values of these ATSs depend on the
structure mentioned.

The following conclusion can be made exclusively on the basis of
the conservation law of the transverse component of the Minkowski
linear momentum. The ATS phenomenon can take place only when both
the reflected and transmitted beams are generated. Independent of
the incident beam's structure, the ATS must be equal zero when one
secondary beam is generated, i.e. when $Q\sp{(\rho)}=1$, and
$Q\sp{(\tau)}=0$, or $Q\sp{(\tau)}=1$, and $Q\sp{(\rho)}=0$; the
former situation takes place when $n\sp{(1)}>n\sp{(2)}$ and
$\theta \sp{(i)}$ exceeds the critical angle for the total
reflection, the latter takes place when the incident beam is
$p$-polarized and $\theta \sp{(i)}$ is equal to the Brewster
angle.

\section*{IV. Conclusions}
\indent

1. We have analyzed the ATSs of the secondary beams on the basis
of conservation laws, which take place in the processes of
reflection and transmission of a wave packet at a plane interface
of two isotropic transparent media. It has been shown that the
values of the ATSs of the reflected and transmitted beams, in case
the shifts exist, are controlled by the conservation law of the
transverse component of the Minkowski linear momentum. Owing to
this law, the angles $\Phi\sp{(\rho )}$ and $\Phi\sp{(\tau )}$
must satisfy the relation (27).

2. The obtained relation (27) is of a general character: by its
derivation, we have assumed that the incident beam is paraxial,
but the distribution of the electromagnetic field inside the beam
has not been specified. This fact is remarkable because the ATS of
every secondary beam depends on this distribution.

3. Without direct calculations of the ATS, only on the basis of
the conservation of the transverse component of the Minkowski
linear momentum, one can conclude that the ATS must be equal zero
when only one secondary beam is generated, i.e. in the
total-reflection and total-transmission regimes.

4. Our analysis was restricted with the processes of reflection
and transmission of the light beam at a single interface of two
isotropic transparent media. However, the conservation of the
transverse component of the Minkowski linear momentum takes also
place when the beam is reflected and transmitted at the stratified
structures of the isotropic transparent media. If in the course of
such a process several well-divided reflected or transmitted beams
are generated (see, for instance, [28]), their ATSs must obey a
condition similar to the relation (27).

\section{Acknowledgment}
\indent

 This work was supported by Estonian Science
Foundation (grant No. 6534)

\end{document}